\documentclass[twocolumn]{aastex631}

\submitjournal{ApJ}

\begin{document}

\title{Hydrogen burning of $^{29}$Si and its impact on presolar stardust grains from classical novae}

\correspondingauthor{Christian Iliadis}
\email{iliadis@unc.edu}

\author{Lori Downen}
\affiliation{Department of Physics \& Astronomy, University of North Carolina at Chapel Hill, NC 27599-3255, USA}
\affiliation{Triangle Universities Nuclear Laboratory (TUNL), Duke University, Durham, North Carolina 27708, USA}

\author[0000-0003-2381-0412]{Christian Iliadis}
\affiliation{Department of Physics \& Astronomy, University of North Carolina at Chapel Hill, NC 27599-3255, USA}
\affiliation{Triangle Universities Nuclear Laboratory (TUNL), Duke University, Durham, North Carolina 27708, USA}

\author{Art Champagne}
\affiliation{Department of Physics \& Astronomy, University of North Carolina at Chapel Hill, NC 27599-3255, USA}
\affiliation{Triangle Universities Nuclear Laboratory (TUNL), Duke University, Durham, North Carolina 27708, USA}

\author{Thomas Clegg}
\affiliation{Department of Physics \& Astronomy, University of North Carolina at Chapel Hill, NC 27599-3255, USA}
\affiliation{Triangle Universities Nuclear Laboratory (TUNL), Duke University, Durham, North Carolina 27708, USA}

\author{Alain Coc}
\affiliation{CNRS/IN2P3, IJCLab, Universit\'e Paris-Saclay, B\^atiment, 104, F-91405 Orsay Campus, France}


\author{Jordi Jos\'{e}}
\affiliation{Departament de F\'\i sica, EEBE, Universitat Polit\`ecnica de Catalunya, c/Eduard Maristany 10, E-08930 Barcelona, Spain}
\affiliation{Institut d'Estudis Espacials de Catalunya, c/Gran Capit\`a 2-4, Ed. Nexus-201, E-08034 Barcelona, Spain}



\begin{abstract}
Presolar stardust grains found in primitive meteorites are believed to retain the isotopic composition of stellar outflows at the time of grain condensation. Therefore, laboratory measurements of their isotopic ratios represent sensitive probes for investigating open questions related to stellar evolution, stellar explosions, nucleosynthesis, mixing mechanisms, dust formation, and galactic chemical evolution. For a few selected presolar grains, classical novae have been discussed as a potential source. For SiC, silicate, and graphite presolar grains, the association is based on the observation of small $N(^{12}$C)/$N(^{13}$C) and $N(^{14}$N)/$N(^{15}$N) number abundance ratios compared to solar values, and abundance excesses in $^{30}$Si relative to $^{29}$Si, as previously predicted by models of classical novae. We report on a direct measurement of the $^{29}$Si(p,$\gamma$)$^{30}$P reaction, which strongly impacts simulated $\delta ^{29}$Si values from classical novae. Our new experimental $^{29}$Si(p,$\gamma$)$^{30}$P thermonuclear reaction rate differs from previous results by up to 50\% in the classical nova temperature range ($T$ $=$ $100$ $-$ $400$~MK), while the rate uncertainty is reduced by up to a factor of $3$. Using our new reaction rate in Monte Carlo reaction network and hydrodynamic simulations of classical novae, we estimate $\delta ^{29}$Si values with much reduced uncertainties. Our results establish $\delta ^{29}$Si values measured in presolar grains as a sensitive probe for assessing their classical nova paternity.
We also demonstrate that $\delta ^{30}$Si values from nova simulations are presently not a useful diagnostic tool unless the large uncertainty of the $^{30}$P(p,$\gamma$)$^{31}$S reaction rate can be significantly reduced.
\end{abstract}

\keywords{Classical Novae (251) --- Explosive nucleosynthesis (503) --- Nuclear reaction cross sections (2087) --- Meteorites (1038)}


\section{Introduction} \label{sec:intro}
Primitive meteorites contain dust grains whose isotopic composition can only be explained by assuming that they condensed in stellar winds or ejecta of exploding stars \citep{Zinner2014,nittler2016}. After formation, these small grains survived the journey through the interstellar medium, for about 1~Gyr, to the local region where the presolar cloud formed about 4.6~Gyr ago. They also survived the homogenization process during the formation of the solar system and were subsequently incorporated into primitive meteorites \citep{Clayton1973}. These so-called presolar stardust grains retain the isotopic composition of the stellar outflows at the time of grain condensation. Therefore, laboratory measurements of their isotopic ratios \citep{Anders1993} represent sensitive probes for investigating open questions related to stellar evolution, stellar explosions, nucleosynthesis, mixing mechanisms, dust formation, and galactic chemical evolution.

Presolar stardust grains identified so far include diamond, SiC, graphite, refractory oxide, silicate, and silicon nitride \citep[see, e.g.,][]{Amari2014}. More than $\approx$90\% of presolar SiC grains are thought to originate from AGB stars, with a small contribution ($\approx$1\%) from type-II supernovae. Among oxide and silicate presolar grains, the estimated relative contributions of AGB stars and type-II supernovae are $\approx$90\% and $\approx$10\%, respectively. Type-II supernovae are believed to be the exclusive source of silicon nitride presolar grains, while they are also the main contributors to graphite presolar grains ($\approx$60\%), compared to a fraction of $\approx$30\% from AGB stars \citep{Hoppe2011}. 

For some presolar grains, classical novae have been discussed as a potential source \citep{Amari2001,Leitner2012,Jose2007,Gyngard2010,Nguyen2014}. Classical novae are caused by the accretion of hydrogen-rich material onto the surface of a white dwarf in a close binary system (see \citet{Jose2016} for a review). Part of the transferred matter accumulates on top of the white dwarf, where it is gradually compressed and heated, until a thermonuclear runaway (TNR) ensues.
As a result, material is ejected into the interstellar medium at high velocities, giving rise to a classical nova. Spectroscopic studies have identified two distinct types. Nova ejecta rich in CNO material point to an underlying CO white dwarf
(``CO novae''), while elemental enrichments in the range of Ne to Ar (besides C, N, O) have been attributed to the presence of an underlying, more massive, ONe white dwarf
(``ONe novae''). The latter novae reach higher peak temperatures and tend to be more energetic than the former \citep{Starrfield1986}.

Observations of classical novae at all wavelengths, ranging from radio waves to $\gamma$-rays \citep{chomiuk2021}, provide important constraints for stellar models as regards to the energetics, mass loss, and shocks associated with these events. Spectroscopically inferred elemental abundances in nova ejecta also provide valuable information, although the abundance estimates are subject to significant uncertainties \citep{jose2008,downen2013}. On the other hand, presolar stardust grains, with their precisely measured isotopic ratios, represent an intriguing probe for nova models. Many classical novae are prolific producers of both carbon-rich and oxygen-rich dust \citep{gehrz1998,gehrz2008} and, therefore, the isotopic composition of the dust grains reflects the hydrodynamical conditions and mixing processes that occur during explosive nuclear burning \citep{starrfield2008}. 

The association of specific presolar grains with a classical novae paternity is based on distinct isotopic abundance signatures. For example, models of ONe novae predict small number abundance ratios of $N(^{12}$C)/$N(^{13}$C) and $N(^{14}$N)/$N(^{15}$N) compared to solar ratios. 
Such models also predict an abundance excess in $^{30}$Si relative to $^{29}$Si \citep{Amari2001,Jose2004,Jose2007}. Since the simulated nova ejecta exhibit more anomalous isotopic ratios compared to the presolar grain measurements, it had frequently been assumed that the presolar grains condensed after the ejecta were diluted with a much larger amount ($\gtrsim$90\%) of close-to-solar matter \citep{Amari2001,Gyngard2010,Leitner2012}. Because neither the mechanism nor the source of this dilution is generally accepted\footnote{Grain condensation will occur between $50$ and $100$ days after the outburst, at a time when part of the ejecta is expected to collide with the accretion disk or the companion star. This process has been suggested as a possible dilution mechanism by \citet{Figueira2018}.}, several authors have recently attempted to match presolar grain compositions with nova model predictions without requiring any (or only a modest amount of) dilution \citep{iliadis2018,Bose2019}. 

Although a nova paternity has been suggested for a number of presolar grains (for a list of nova candidate grains see Table~2 in \citet{iliadis2018} or Table~1 in \citet{Bose2019}), the issue is intensely debated because counter-arguments favor a supernova origin for many of these grains \citep{Nittler_2005,Liu2016}. Improved stellar model predictions of isotopic signatures, including for silicon, are highly desirable to shed light onto the origin of these presolar grains. 

\begin{figure}[hbt!]
\centering
\includegraphics[width=0.95\linewidth]{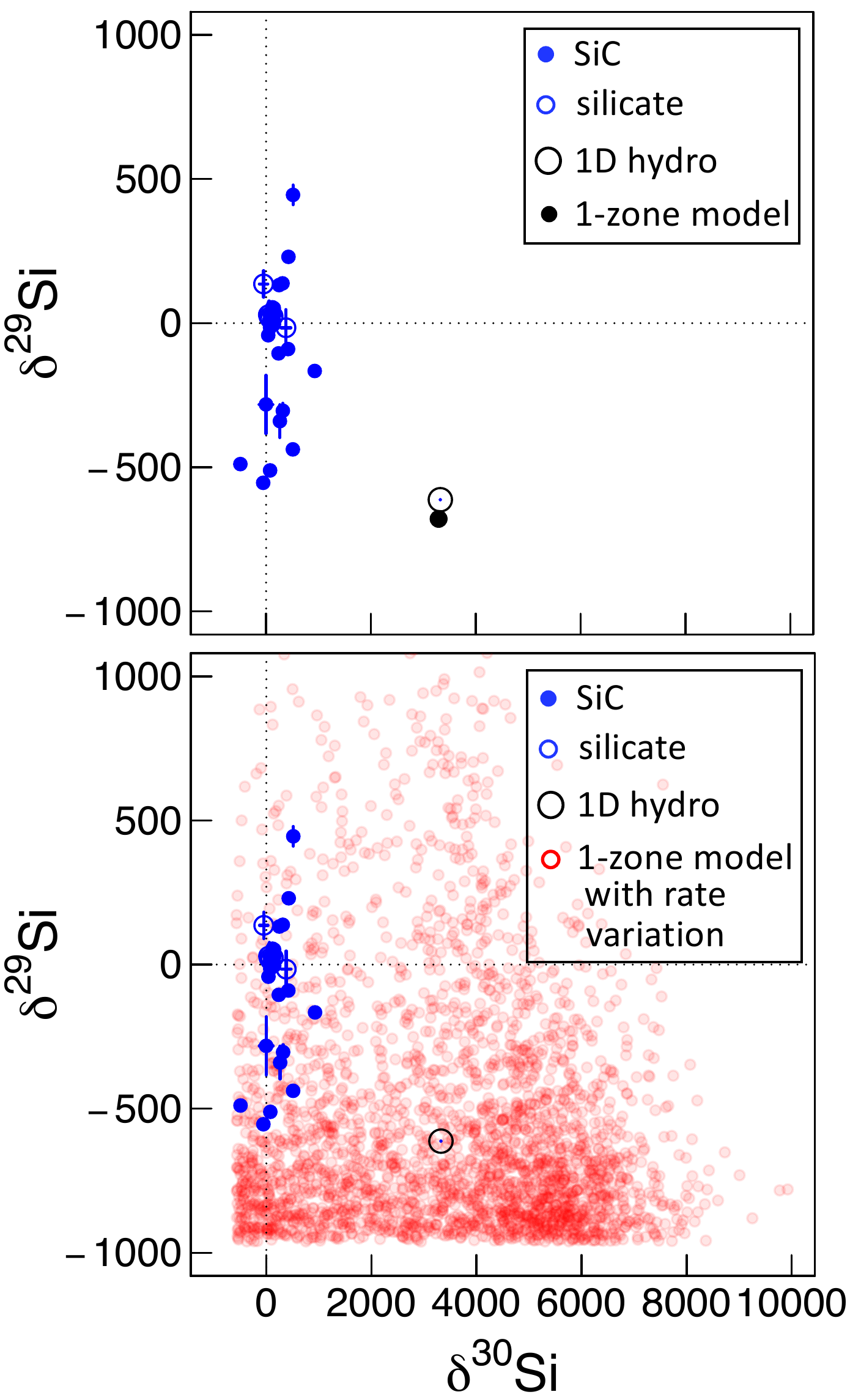}
\caption{Silicon isotopic ratios in SiC (solid blue circles) and silicate (open blue circles) presolar nova candidate grains. For the source of grain data, see Tab.~2 in \citet{iliadis2018} or Tab.~1 in \citet{Bose2019}.  The data are plotted as $\delta$ values, which represent deviations from solar ratios in parts per thousand. The ratios simulated using a 1D hydrodynamic model \citep[model ONe2 of][]{jose2020} are depicted as an open black circle. (Top) The solid black circle represents the results using a parametric one-zone nucleosynthesis calculation (Section~\ref{sec:impact}), assuming median rates for all nuclear reactions in the network, i.e., without consideration of rate uncertainties. (Bottom) Model ratios obtained with the previous $^{29}$Si(p,$\gamma$)$^{30}$P reaction rate from 5,000 one-zone models (open red circles), where for each run the thermonuclear rates of all reactions in the network were independently sampled within their uncertainties. Notice that about 8\% of the 5,000 simulations result in solutions located in the upper right quadrant.
}
\label{fig:fu3}
\end{figure}

In the top panel of Figure~\ref{fig:fu3} we show silicon isotopic ratios as $\delta$ values, e.g., $\delta^i Si$ $\equiv$ $\delta^i Si/^{28}Si$ $\equiv$ $[(^i Si/^{28}Si)_{grain}/(^i Si/^{28}Si)_{solar}$ $-$ $1$] $\times$ $1000$, measured in presolar nova candidate grains (SiC: solid blue circles; silicate: open blue circles). 
The black open circle depicts simulated ratios, mass-weighted over the entire ejected envelope, that have been obtained using the one-dimensional hydrodynamic code SHIVA \citep{Jose2016} assuming a ONe model with a white dwarf mass of $1.25$M$_\odot$ (model ONe2\footnote{We have chosen model ONe2 as an initial example because it was computed with the conventional prescription of mixing white dwarf with solar matter, i.e., a pre-enriched composition at the start of the simulation; by contrast, model ONe1 was computed under a new methodology, where the time-dependent amount of mass dredged-up from the outer white dwarf layers, and the time-dependent convective velocity profile throughout the envelope, are extracted from 3D simulations and subsequently implemented into the 1D code SHIVA to complete the simulation through the late expansion and ejection stages of the nova outburst. Furthermore, we have chosen a ONe model instead of a CO model because little silicon nucleosynthesis occurs in the latter except for the highest CO white dwarf masses, as will be shown later.} in \citet{jose2020}). The simulation was performed with the median thermonuclear reaction rates listed in the STARLIB library\footnote{The STARLIB library can be found at \url{https://starlib.github.io/Rate-Library/}.} \citep{Sallaska}. Any post-explosion dilution of the simulated abundances with solar-like matter will give rise to a composition that, depending on the degree of dilution, is located on a straight line connecting the open black circle with the origin ($\delta^{29} Si$ $=$ $\delta^{30} Si$ $=$ $0$). Since the diluted composition can only explain the grains of the lower right quadrant, it is frequently argued that the grains of the upper right quadrant most likely originate from another source, such as type-II supernovae (see, e.g., the arguments put forward by \citet{Liu2016} as regards to ``C2 grains'').

The isotopic signatures discussed above are the result of nuclear reactions taking place during the TNR. Since the thermonuclear reaction rates are subject to uncertainties, it is important to investigate their impact on the simulations. Discussions of reaction rate uncertainties related to classical novae can be found in \citet{Jose2001,iliadis2002,Jose2004,vanRaai2008}, but are nearly absent in the more recent literature on nova candidate grains. 

In Section~\ref{sec:impact}, we will demonstrate the strong impact of current reaction rate uncertainties on the isotopic composition in nova ejecta. In Section~\ref{sec:exp}, we report on a direct measurement of the $^{29}$Si(p,$\gamma$)$^{30}$P reaction at energies relevant to nova nucleosynthesis. The impact of our new reaction rate is investigated in Section~\ref{sec:simu}. A concluding summary is given in Section~\ref{sec:summary}.

\section{Impact of thermonuclear reaction rate uncertainties} \label{sec:impact}
We performed a one-zone reaction network simulation by assuming an analytical parameterization for the thermodynamic trajectories of the explosion, $T(t)$ $=$ $T_{peak} e^{-t/\tau_T}$ and $\rho(t)$ $=$ $\rho_{peak} e^{-t/\tau_\rho}$, where $t$ $\ge$ $0$ is the time since peak temperature, $T_{peak}$, or peak density, $\rho_{peak}$. The quantities $\tau_T$ and $\tau_\rho$ are the times at which temperature and density, respectively, have fallen to $1/e$ of their peak values. We will use values of $T_{peak}$ $=$ $238$~MK and $\rho_{peak}$ $=$ $258$~g~cm$^{-3}$, which are identical with the peak temperature and the density at peak temperature achieved in model ONe2 of \citet{jose2020}. For the expansion time scales we choose values of $\tau_T$ $=$ $3000$~s and $\tau_\rho$ $=$ $50$~s. Thermonuclear reaction rates are strongly temperature-dependent. Therefore, most of silicon nucleosynthesis takes place near peak temperature and the simulated abundances are not very sensitive to the expansion time scales \citep[see also][]{iliadis2018}. For the initial composition we adopted the same as \citet{jose2020}, i.e., a mixture of 23\% white dwarf matter with 77\% solar-like matter, where the ONe white dwarf composition was adopted from \citet{ritossa1996}. The resulting silicon isotopic ratios are depicted by the black solid circle in the top panel of Figure~\ref{fig:fu3}. As can be seen, the result of the parametric one-zone calculation is close to that of the hydrodynamic model (open black circle). 

The advantage of using a parametric model is that we can perform a series of many network calculations by sampling the reaction rate uncertainties independently\footnote{The rate of each nuclear reaction in the network is uncorrelated from the rate of any other nuclear reaction. The only exceptions are pairs of forward and corresponding reverse reactions. These are highly correlated because they must always be multiplied by the same rate variation factor. Because all rates (except for the reverse ones) are sampled independently and the silicon isotopic abundance ratios are obtained at each Monte Carlo step, correlations among reactions and abundances are fully taken into account.} for all nuclear reactions in the network. For these simulations, we also use the STARLIB reaction rate library. Apart from providing median thermonuclear rates for all reactions of interest, it also lists rate probability density functions on a grid of temperatures between 1~MK and 10~GK \citep{Sallaska}. This information is valuable because it can be used to randomly sample the rate of each nuclear reaction in the network according to its individual rate probability density. 

The bottom panel of Figure~\ref{fig:fu3} shows the result of this procedure for a series of 5,000 network calculations, where each red circle corresponds to the silicon isotopic ratios of a single network run. The total number of runs was sufficient to quantify the dispersion in the results. The average of the Monte Carlo samples is consistent with the black open (``1D hydro'') or full (``1-zone model'') circles. Solutions marked by the red circles represent the composition of pure ejecta, i.e., we have not assumed any dilution. It can be seen that, once thermonuclear reaction rate uncertainties are considered, the simulated silicon abundance ratios scatter over a significant region of parameter space. This large scatter has not been noticed before. Since many of the red circles are located in a region consistent with an excess in $^{29}$Si relative to $^{30}$Si, our results do not support the general assumption made frequently in the literature that nova models predict an abundance excess in $^{30}$Si relative to $^{29}$Si (Section~\ref{sec:intro}). Many plausible solutions (about 8\% among 5,000) are even located in the upper right quadrant, e.g., the region of the C2 grains \citep{Liu2016}, which cannot be reached when performing the simulations with the median thermonuclear rate for each reaction in the network, even if dilution is considered. To summarize, when current reaction rate uncertainties are taken into account, it is not possible to assess if nova models produce over- or underabundances of $^{29}$Si or $^{30}$Si relative to solar, or if any abundance anomaly is produced at all.

As will be seen below, the large scatter observed in Figure~\ref{fig:fu3} is mainly caused by current rate uncertainties of just two nuclear reactions: the $^{29}$Si(p,$\gamma$)$^{30}$P reaction rate determines the $^{29}$Si-to-$^{28}$Si abundance ratio, while the $^{30}$P(p,$\gamma$)$^{31}$S rate impacts the $^{30}$Si-to-$^{28}$Si ratio. For the first reaction, we assumed a rate uncertainty factor of $3$, which is equal to the maximum uncertainty reported in the evaluation of \citet{ILIADIS2010b} at the nova temperature range of $T$ $\approx$ $100$ $-$ $400$~MK. For the second reaction, the rate uncertainty adopted in STARLIB is a factor of $10$. 

\section{Experimental procedure and results} \label{sec:exp}
Current uncertainties in the $^{29}$Si(p,$\gamma$)$^{30}$P thermonuclear reaction rate are caused by known low-energy resonances, which were previously measured with limited detection sensitivity, and by contributions from yet undetected resonances near the proton threshold. Recently, \citet{Lotay2020} estimated the $^{29}$Si $+$ $p$ reaction rate indirectly, based on experimental results for nuclear levels in $^{30}$P, and they suggested a direct measurement of the reaction,  $^{29}$Si(p,$\gamma$)$^{30}$P, that actually takes place in a classical nova.

We measured the $^{29}$Si(p,$\gamma$)$^{30}$P reaction directly at the Triangle Universities Nuclear Laboratory (TUNL) using the two ion accelerators at the Laboratory for Experimental Nuclear Astrophysics (LENA). Resonances above $300$~keV bombarding energy were measured using a 1-MV model JN Van de Graaff accelerator, which provided proton beam intensities up to $40$~$\mu$A on target. Measurements below $250$~keV bombarding energy were carried out with the Electron Cyclotron Resonance Ion Source (ECRIS) accelerator. The maximum proton beam current on target was 2.1~mA. 

Gamma rays emitted from the target were analyzed with a $135$\% relative efficiency coaxial High-Purity Germanium (HPGe) detector surrounded by a $16$-segment NaI(Tl) annulus. These detectors, which comprise the LENA $\gamma\gamma$-coincidence spectrometer \citep{longland2006,buckner2015},
are capable of reducing the room background in the energy region below $2.6$~MeV by large factors. The spectrometer is well characterized \citep{carson2010}, allowing for the determination of reliable singles and coincidence detection efficiencies in conjunction with Geant4 (Monte Carlo transport) simulations \citep{howard2013}. The measured pulse height spectra were modeled using a binned likelihood method with Monte Carlo simulated spectra \citep{dermigny16}. The fraction of the experimental spectrum belonging to each template was obtained using a Bayesian statistical approach. This allowed for the determination of the primary $\gamma$-ray branching ratios and the total number of $^{29}$Si(p,$\gamma$)$^{30}$P reactions taking place during the experiment. 
%

\begin{figure*}[hbt!]
\centering
\includegraphics[width=1.0\linewidth]{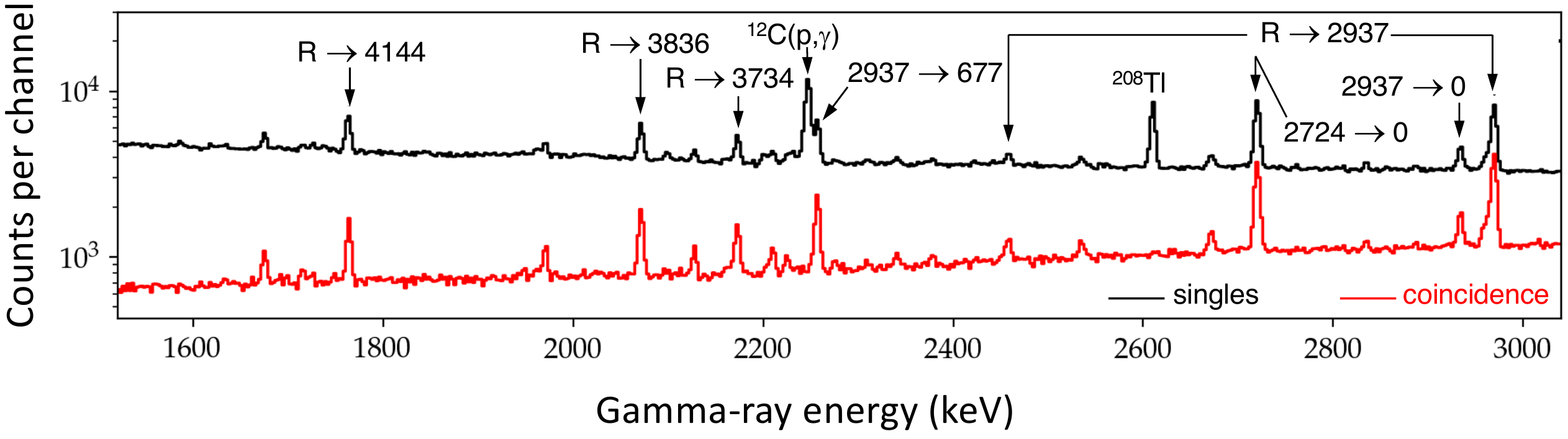}
\caption{Singles (black) and coincidence (red) pulse-height spectra measured at the $E_r^{cm}$ $=$ $314$~keV resonance in $^{29}$Si(p,$\gamma$)$^{30}$P, which is most important for $^{29}$Si nucleosynthesis in classical novae. The strongest peaks are labeled according to their origin. The displayed energy range exhibits three weak, previously unobserved $\gamma$-ray transitions, labeled as ``$R \rightarrow 3734$'', ``$R \rightarrow 3836$'', and ``$R \rightarrow 4144$'', where the symbol $R$ denotes the excitation energy, $E_x$ $=$ $5908$~keV, of the corresponding level in the $^{30}$P compound nucleus. The value behind the arrow stands for the excitation energy (in keV) of the $^{30}$P level populated in the transition. Notice the overall reduction of the background in the coincidence spectrum. Both the environmental background peak at $2614$~keV (from $^{208}$Tl decay) and the beam-induced background peak (from $^{12}$C(p,$\gamma$)$^{13}$N) are absent in the coincidence spectrum. 
}
\label{fig:spectrum}
\end{figure*}

The $^{29}$Si target was implanted using the Source of Negative Ions by Cesium Sputtering (SNICS) at the Centre de Spectrom\'etrie Nucl\'eaire et de Spectrom\'etrie de Masse (CSNSM) in Orsay, France. The $^{29}$Si$^-$ beam was produced from natural silicon metalloid and implanted at $80$~keV bombarding energy into a $0.5$-mm thick tantalum sheet. Prior to implantation, the tantalum backing was chemically etched and then outgassed in high vacuum by resistive heating to remove contaminants. 
The target thickness was found to be $11.4 \pm 0.3$~keV near $\approx$400~keV bombarding energy, while the stoichiometry of the target layer was Ta/$^{29}$Si $=$ $1.2 \pm 0.2$. Resonance yield curves measured during the course of the experiment demonstrated that both the maximum yield and thickness of the target were unchanged after an accumulated proton charge of $17$~C.

We measured energies and resonance strengths\footnote{The resonance strength, which represents the integrated nuclear reaction cross section of a narrow resonance, is defined for the $^{29}$Si(p,$\gamma$)$^{30}$P reaction by $\omega\gamma$ $\equiv$ $ (2J+1) \Gamma_p \Gamma_{\gamma}/(4\Gamma$), with $\Gamma_p$, $\Gamma_{\gamma}$, $\Gamma$, and $J$ the proton width, $\gamma$-ray width, total width, and spin of the resonance, respectively.} of three $^{29}$Si(p,$\gamma$)$^{30}$P resonances, at center-of-mass energies of $E_r^{cm}$ $=$ $303$~keV, $314$~keV, and $403$~keV. We also determined an experimental upper limit for the strength of a fourth resonance, at $E_r^{cm}$ $=$ $215$~keV. Details on our analysis will be provided in a forthcoming publication. Here we will summarize the results. 

The $E_r^{cm}$ $=$ $403$~keV resonance is well-known and has been measured previously several times \citep{Reinecke1985,RIIHONEN1979251}. We used its recommended resonance strength, $\omega\gamma$ $=$ $0.220 \pm 0.025$~eV \citep{SARGOOD198261}, as a standard for determining the strengths, or upper limits, of other resonances measured in the present work. 

The $E_r^{cm}$ $=$ $314$~keV resonance, which is most important for $^{29}$Si nucleosynthesis in classical novae, has been measured previously \citep{PhysRev.110.96,Harris1969,Reinecke1985}, but the reported resonance strength values differ by large factors and have large uncertainties. Our result for the resonance strength is $\omega\gamma$ $=$ $0.0207 \pm 0.0027$~eV, representing a $13$\% uncertainty. Most of the uncertainty is caused by that of the standard resonance at $E_r^{cm}$ $=$ $403$~keV ($11$\%). Other sources of systematic uncertainty derive from the Geant4 simulations ($1.3$\%), beam charge integration ($3.0$\%), stopping powers ($5.0$\%), and the detector geometry ($5.0$\%). The statistical uncertainty is less than $1.0$\%. Our result has a much smaller uncertainty than all previously reported values. Sections of singles and coincidence pulse-height spectra in the region of three previously unobserved, weak primary $\gamma$-ray transitions (labeled as ``$R \rightarrow 3734$'', ``$R \rightarrow 3836$'', and ``$R \rightarrow 4144$'') are displayed in Figure~\ref{fig:spectrum}, demonstrating the significant background reduction in the coincidence spectrum (see figure caption for details).  

The resonance at $E_r^{cm}$ $=$ $303$~keV had not been detected previously. In total, we observed five different primary transitions, both in the singles and the coincidence spectra. Our precisely measured value for the resonance energy is $E_r^{cm}$ $=$ $303.4 \pm 1.0$~keV. For the resonance strength, we find $\omega\gamma$ $=$ $(8.8 \pm 1.5) \times 10^{-5}$~eV, representing a $17$\% uncertainty. In previous rate evaluations \citep{LONGLAND2010,ILIADIS2010b,ILIADIS2010c,iliadis2010d}, this resonance had an energy of $E_{r,2010}^{cm}$ $=$ $296 \pm 12$~keV and a strength of $\omega\gamma_{2010}$ $\approx$ $4 \times 10^{-5}$~eV. The latter order-of-magnitude estimate had been obtained indirectly using nuclear structure information. 

We also searched for a resonance at $E_r^{cm}$ $=$ $215$~keV, which is expected based on the known level structure of the $^{30}$P compound nucleus \citep{ensdf2021}. Singles and coincidence spectra were accumulated for a beam charge of $\approx$10~C. While we did not observe any primary or secondary transitions in $^{30}$P, we measured for the resonance strength an upper limit of $\omega\gamma$ $\le$ $3.3 \times 10^{-7}$~eV (97.5\% coverage probability).

\begin{figure}[hbt!]
\centering
\includegraphics[width=0.95\linewidth]{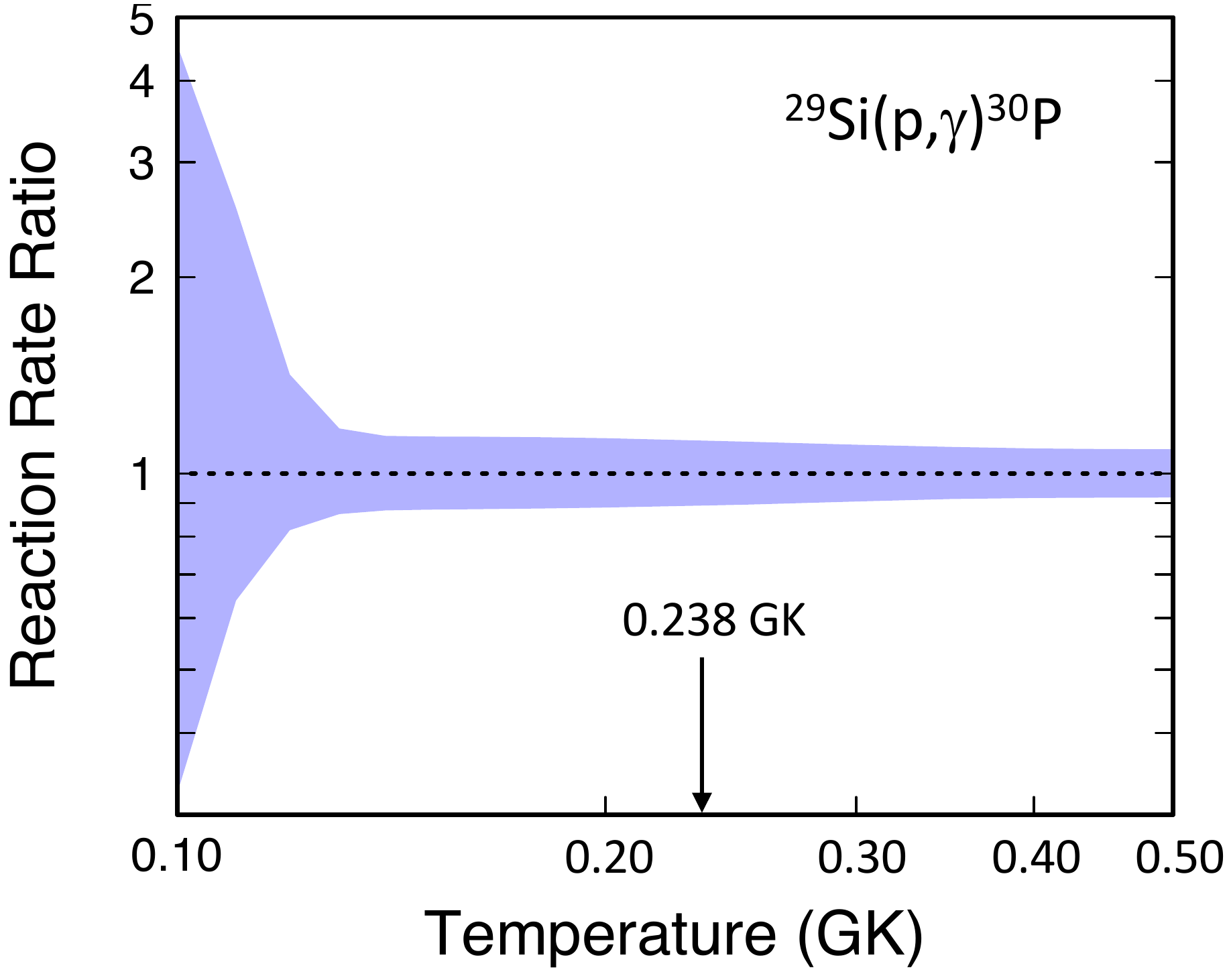}
\caption{Present $^{29}$Si(p,$\gamma$)$^{30}$P thermonuclear reaction rate uncertainties in the classical nova temperature region. The upper and lower bounds of the blue region represent the 16 and 84 percentiles of the total reaction rate probability density. Because reaction rates vary over many orders of magnitude, we display the results as ratios to the median (50 percentile) of the total rate. The peak temperature achieved in model ONe2 of \citet{jose2020} is indicated by the arrow. Near this temperature region, the uncertainty of the new rate amounts to only 12\%. 
}
\label{fig:rates}
\end{figure}

Based on our new experimental results, together with our evaluation of the level structure near the proton threshold in $^{30}$P, we derived a new $^{29}$Si(p,$\gamma$)$^{30}$P thermonuclear reaction rate with significantly smaller uncertainties at nova temperatures compared to previous results. We estimated the total rate using the Monte Carlo procedure presented in \citet{LONGLAND2010}, which fully implements the uncertainties of all experimental input quantities (e.g., resonance energies, strengths, partial widths, non-resonant S-factors). Our reaction rates are displayed in Figure~\ref{fig:rates} versus the temperature region most important for classical novae ($T$ $\approx$ $100$ $-$ $400$~MK). The peak temperature achieved in model ONe2 of \citet{jose2020} is indicated by the arrow, which is near the region where most of the silicon nucleosynthesis takes place in this particular model. At this temperature, our new rate is $\approx$50\% larger than the previous estimate \citep{ILIADIS2010b}, while the rate uncertainty has been reduced by almost a factor of $3$. Notice that the rate uncertainty starts to increase below 130~MK because of yet unobserved resonances. However, these low temperatures are not important for silicon nucleosynthesis in classical novae.

\section{Consequences for nova candidate grains} \label{sec:simu}
\begin{figure}[hbt!]
\centering
\includegraphics[width=0.95\linewidth]{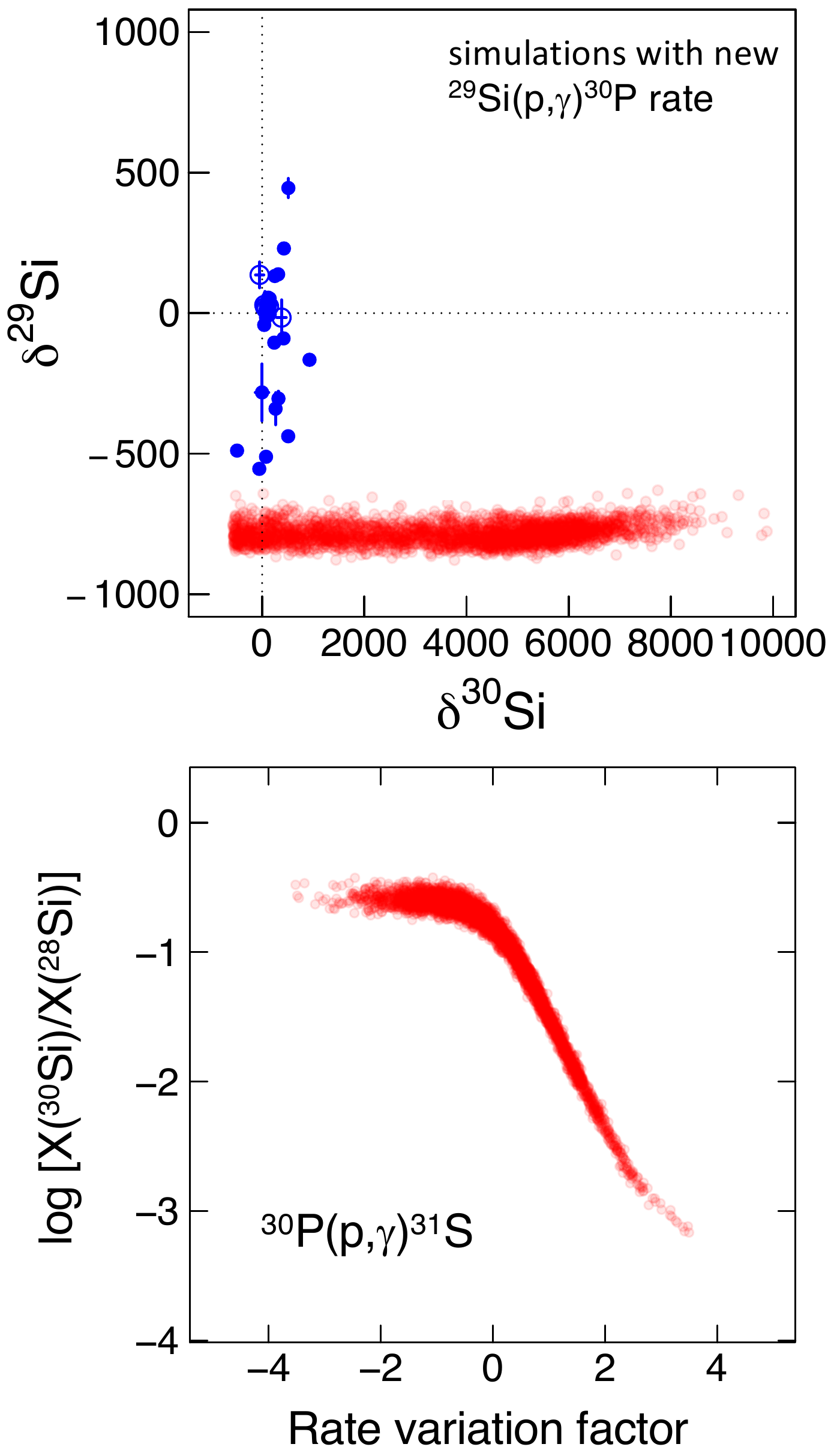}
\caption{(Top) Same as Figure~\ref{fig:fu3}, but using our new $^{29}$Si(p,$\gamma$)$^{30}$P thermonuclear reaction rate in the rate sampling of all reactions in the network. Comparison to the bottom panel of Figure~\ref{fig:fu3} reveals the drastic reduction in the scatter of the $^{29}$Si abundance prediction. (Bottom) The $^{30}$Si-to-$^{28}$Si mass fraction ratio versus rate variation factor of $^{30}$P(p,$\gamma$)$^{31}$S for the same Monte Carlo calculation as displayed in the top panel. Notice the strong correlation, which causes the large scatter of the predicted $^{30}$Si abundances evident in the top panel of this figure and the bottom panel of Figure~\ref{fig:fu3}.
}
\label{fig:new}
\end{figure}

We repeated the Monte Carlo reaction network simulations (Section~\ref{sec:impact}), with the only change of using our new $^{29}$Si(p,$\gamma$)$^{30}$P thermonuclear reaction rate and its uncertainty in the rate sampling of all reactions in the network. The results are shown in the top panel of Figure~\ref{fig:new}. Comparison to the bottom panel of Figure~\ref{fig:fu3} demonstrates the drastic reduction in the uncertainty of the predicted $^{29}$Si abundances resulting from the present experiment. We find that 
the 12\% uncertainty in the $^{29}$Si(p,$\gamma$)$^{30}$P rate at the achieved peak temperature translates into an about equal uncertainty in the simulated value of $\delta ^{29}$Si.

\begin{figure}[hbt!]
\centering
\includegraphics[width=1.0\linewidth]{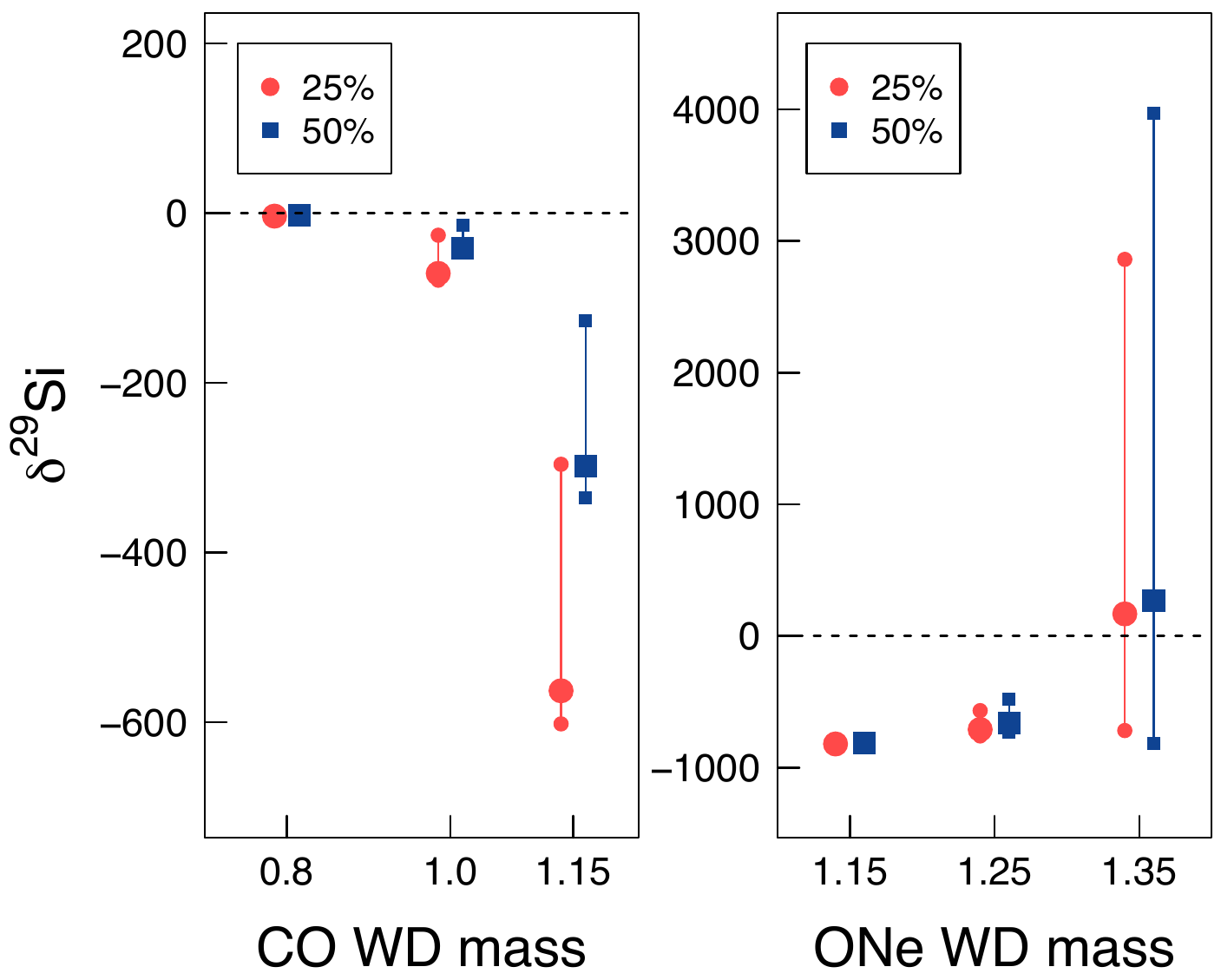}
\caption{Estimated $\delta ^{29}$Si values from one-dimensional hydrodynamic CO nova (left panel) and ONe nova (right panel) models versus the  mass of the underlying white dwarf (in units of the solar mass). The red circles and blue squares correspond to an initial (pre-explosion) composition resulting from a 25\% and 50\% admixture, respectively, of outer white dwarf matter. For each nova model, the vertical line indicates the range of values obtained across all ejected shells, with the mass-weighted average value over the entire ejected envelope indicated by the large central circle or square. Note that for some models the variation of $\delta ^{29}$Si is so small that only the latter symbols are visible. The horizontal dashed lines correspond to the initial pre-mixed composition prior to the TNR. These results are obtained with median reactions rates only, i.e., they disregard rate variations. However, the effect of rate uncertainties on the displayed values is very small (see text).
}
\label{fig:cn}
\end{figure}

To quantify the range of $\delta ^{29}$Si values obtained in classical nova models, we performed a series of one-dimensional hydrodynamic simulations using the code SHIVA \citep{Jose2016}. In total, $12$ models were computed, covering a range of CO and ONe white dwarf masses and 
pre-mixed compositions. Again, thermonuclear reaction rates were adopted from the STARLIB library, except that the present rate was used for the $^{29}$Si(p,$\gamma$)$^{30}$P reaction. The results are presented in Figure~\ref{fig:cn}. The left and right panel depicts $\delta ^{29}$Si values for CO and ONe novae, respectively. The red and blue symbols refer to pre-mixed compositions involving 25\% and 50\% of white dwarf matter, respectively. The vertical lines indicate the range of values obtained across all ejected shells, while the larger central symbols on each vertical line correspond to the mass average over the entire ejected envelope. The horizontal dashed lines ($\delta ^{29}$Si $=$ $0$) represent the values of the pre-mixed composition prior to the TNR. It can be seen that the explosion leaves the $^{29}$Si abundance for the 0.8M$_{\odot}$ and 1.0M$_{\odot}$ CO models (left panel) nearly unchanged because the achieved peak temperatures are relatively low. For the 1.15M$_{\odot}$ CO model, we find ranges of $-600$ $\le$ $\delta ^{29}$Si $\le$ $-300$ (25\% mixing) and $-340$ $\le$ $\delta ^{29}$Si $\le$ $-130$ (50\% mixing). The 1.15M$_{\odot}$ and 1.25M$_{\odot}$ ONe models predict values of $\delta ^{29}$Si $\approx$ $-810$ and $-600$, respectively, which are nearly independent of the initial composition. For the 1.35M$_{\odot}$ ONe model we find wide ranges of $-720$ $\le$ $\delta ^{29}$Si $\le$ $+2900$ (25\% mixing) and $-810$ $\le$ $\delta ^{29}$Si $\le$ $+4000$ (50\% mixing). Note, that the 1.35M$_{\odot}$ ONe model is the only one we computed that predicts a net production of $^{29}$Si (i.e., positive $\delta ^{29}$Si values) in specific ejected shells. All of these results have been obtained by adopting median reaction rates only. We performed additional tests, which showed that the results presented in Figure~\ref{fig:cn} are robust, in the sense that the additional variation introduced by current reaction rate uncertainties is very small ($\approx$ 12\%). Therefore, our results establish $\delta ^{29}$Si values measured in presolar stardust grains as an important diagnostic tool for assessing the classical nova paternity.

Our results can also be compared with previous estimates of silicon isotopic ratios from hydrodynamic models of classical novae. The values listed in Table~3 of \citep{Jose2004} have frequently been used to interpret presolar grain measurements. For the models involving the smallest white dwarf masses (i.e., 0.8M$_{\odot}$ and 1.0M$_{\odot}$ for CO novae; 1.15M$_{\odot}$ and 1.25M$_{\odot}$ for ONe novae), previous and present $\delta ^{29}$Si values agree approximately. For the 1.15M$_{\odot}$ CO nova model, the simulated values differ by about a factor of $2$, both across individual ejected shells and their mass-weighted average over the entire ejected envelope. The largest deviations between present and previous results are found for the 1.35M$_{\odot}$ ONe nova model pre-enriched with 50\% white dwarf matter. In this case, the previous mass-weighted average was $\delta ^{29}$Si $=$ $-60$, while the result with our new $^{29}$Si(p,$\gamma$)$^{30}$P rate is $\delta ^{29}$Si $=$ $+270$. Consequently, contrary to previous practice, the condition $\delta ^{29}$Si $<$ $0$ should not be applied for claiming a classical nova grain paternity. We emphasize that the previous results \citep{Jose2004} did not take any reaction rate uncertainties into account, which were substantial prior to our experiment, as pointed out above.


%

While our experiment reduced the scatter in the $^{29}$Si abundance prediction significantly, the large scatter in the simulated $^{30}$Si abundance remains. This scatter is mainly caused by current uncertainties in the $^{30}$P(p,$\gamma$)$^{31}$S reaction rate. The impact of this rate on nova nucleosynthesis has been pointed out by \citet{Jose2001,iliadis2002,2014AIPA....4d1004W}. The bottom panel of Figure~\ref{fig:new} depicts the strong correlation of the $^{30}$Si-to-$^{28}$Si mass fraction ratio and the $^{30}$P(p,$\gamma$)$^{31}$S rate variation factor\footnote{We define the ``rate variation factor'', $p$, of a specific nuclear reaction as in \citet{Iliadis2015}, i.e., for each Monte Carlo network run a sampled rate is modified from its median value by a factor $(f.u.)^p$, where $f.u.$ is the factor uncertainty provided in STARLIB together with the median rate.} for the same Monte Carlo simulation as displayed in the top panel. Although nuclear structure information relevant for this reaction has been obtained experimentally \citep{Parikh2011,Irvine2013,PhysRevLett.116.102502,kankainen2017,PhysRevC.102.045806} and theoretically \citep{Brown2014}, large uncertainties remain in its thermonuclear rate. Therefore, $\delta ^{30}$Si values are not a useful probe of a nova paternity, unless the $^{30}$P(p,$\gamma$)$^{31}$S reaction rate can be improved significantly.

\section{Summary} \label{sec:summary}
We reported on a direct measurement of the $^{29}$Si(p,$\gamma$)$^{30}$P reaction. The experiment was performed with exceptionally high proton beam currents and a novel $\gamma$-ray coincidence spectrometer, resulting in a significantly improved signal-to noise ratio compared to previous work. We measured the energies and strengths of three low-energy resonances, located at center-of-mass energies of $E_r^{cm}$ $=$ $303$~keV, $314$~keV, and $403$~keV, and determined an experimental upper limit for the strength of a fourth resonance, located at $E_r^{cm}$ $=$ $215$~keV. Based on these results, we presented a new $^{29}$Si(p,$\gamma$)$^{30}$P thermonuclear reaction rate. The new median rate differs from previous results by up to 50\% in the classical nova temperature range ($T$ $=$ $100$ $-$ $400$~MK), while the rate uncertainty has been reduced by up to a factor of $3$. 

We performed Monte Carlo reaction network and hydrodynamic simulations to demonstrate the strong impact of the $^{29}$Si(p,$\gamma$)$^{30}$P reaction rate on the $^{29}$Si-to-$^{28}$Si abundance ratio obtained from classical nova models. Using the previous rate, the simulated values of $\delta ^{29}$Si scattered over a large range, between $-1000$ and $+1000$, with about 8\% of all solutions resulting in positive $\delta ^{29}$Si values. Previously, it was not possible to assess if, for example, ONe nova models produce over- or underabundances of $^{29}$Si or $^{30}$Si relative to solar, or if any silicon abundance anomaly is produced at all. Using instead our new reaction rate, the uncertainty of the simulated $\delta ^{29}$Si values is significantly reduced. Therefore, our results establish experimental $\delta ^{29}$Si values as an important diagnostic tool for assessing the classical nova paternity of presolar stardust grains. However, significant uncertainties remain in the simulated $\delta ^{30}$Si values, which precludes their use for assessing a classical nova paternity. The simulated $^{30}$Si abundance is strongly influenced by the $^{30}$P(p,$\gamma$)$^{31}$S reaction, for which improved rate estimates are highly desirable.


\begin{acknowledgments}
The authors would like to thank Udo Friman-Gayer for helpful comments. This work was supported in part by the DOE, Office of Science, Office of Nuclear Physics, under grants DE-FG02-97ER41041 (UNC) and DE-FG02-97ER41033 (TUNL). We also acknowledge partial support by the Spanish MINECO grant PID2020-117252GB-I00, by ChETEC-INFRA (EU project no. 101008324), and by the EU FEDER fund.
\end{acknowledgments}

\bibliography{ms.bib}{}
\bibliographystyle{aasjournal}



\end{document}